\documentclass[twocolumn,twoside,slac_two]{revtex4}
\usepackage{graphicx}
\usepackage{fancyhdr}
\usepackage{xspace}
\newcommand{\ke}{\ensuremath{K^\pm\to e^\pm\bar{\nu_e}}\xspace}
\newcommand{\km}{\ensuremath{K^\pm\to \mu^\pm\bar{\nu_\mu}}\xspace}
\def\Journal#1#2#3#4{{\it #1} {\bf #2}, #3 (#4)}

\begin{document}

\title{$K\to e\nu$ decays and lepton flavor violation searches with Kaons}

%

\author{T. Spadaro}
\affiliation{Laboratori Nazionali di Frascati dell'INFN, Frascati, Italy}

\begin{abstract}
Recent Kaon decay studies seeking lepton-flavor violating (LFV) new-physics
effects are briefly discussed. 
The main focus is set on precise measurements of rare or not-so-rare decays
aiming at finding evidence of deviations from the SM prediction rather than on 
the results from direct searches of LFV transitions forbidden or ultra-rare
in the SM.
\end{abstract}

\maketitle

\thispagestyle{fancy}


\section{Introduction}
A significant effort has been devoted along the years to isolate signals from
lepton flavor violating (LFV) transitions, which are forbidden or ultra-rare
in the Standard Model (SM). The sensitivity to decays such as $\mu\to e\gamma$,
$\mu\to eee$, $K_L\to \mu e(+\pi^0\mbox{'s})$, and others roughly improved by
two orders of magnitude for each decade~\cite{landsbergf04}. The activity in
this field is still alive, with results coming fairly 
recently~\cite{ktev+08:klpi0mue}. No signal has been observed, thus ruling out
SM extensions with LFV amplitudes with mediator masses below $\sim100$~TeV.  

These results allowed the focus to be put on the detection of NP-LFV effects 
in loop amplitudes, by studying specific processes suppressed in the SM.
Kaon physics is very well suited to match the precisions required for this
task, the experiments benefiting from low background level and the theoretical
predictions carefully accounting for radiative corrections to reach uncertainties of
the level of 0.1\% or better.

In the SM, electrons and muons differ only by mass and coupling to the Higgs. This allows
the seeking of deviations from prediction in semileptonic ($K_{l3}$)  and leptonic ($K_{l2}$) kaon decays. 

Precise measurements of the semileptonic decay widths have been performed in recent years to extract
the $V_{us}$ parameter of the CKM matrix. 
The following expression is used:
\begin{eqnarray*}\nonumber
\Gamma^i(K_{e3(\gamma),\,\mu3(\gamma)})&=&|V_{us}|^2\frac{C_i^2 G^2 M^5}{128\pi^3} S_{\rm
EW}\: \\ \nonumber & & |f^{K^0}_+(0)|^2 I^{i}_{e3,\,\mu3}\: 
  (1+\delta_{e3,\,\mu3}^{i}),\\ 
\end{eqnarray*}
where $i$ indexes $K^0\to\pi^-$ and $K^+\to\pi^0$ transitions for which $C_i^2 =1$ and 1/2, 
respectively, $G$ is the Fermi
constant, $M$ is the appropriate kaon mass, and $S_{\rm EW}$ is a
universal short-distance electroweak correction~\cite{as}. 
The $\delta^i$ term accounts for long-distance radiative corrections depending on the 
meson charges and lepton masses and, for $K^\pm$,
for isospin-breaking effects. These corrections are presently known at the few-per-mil level~\cite{aa}.
The $f^{K^0}_+(0)$ form factor parametrizes the vector-current transition 
$K^0\to\pi^-$ at zero momentum transfer $t$, while the dependence of
vector and scalar form factors on $t$ enter into the determination of the integrals $I_{e3,\,\mu3}$
of the Dalitz-plot density over the physical region. Four experiments with different techniques 
provided a new comprehensive set of measurements for all of the quantities appearing in the above equation.
Results have been obtained for $K_L$ (KLOE~\cite{KLOE+06:KLBR,KLOE+06:KLe3FF,KLOE+05:KLlife}, 
KTeV~\cite{KTeV+04:KLBR}, 
NA48~\cite{NA48+04:KLBR,NA48+04:KL3p0KS2p0}), 
$K^\pm$ (KLOE~\cite{KLOE+08:KPlife,KLOE+08:KPBR}, NA48~\cite{NA48/2+07:KPBR}), $K^-$ (ISTRA+~\cite{ISTRA+07:KPL3}), 
and $K_S$ (KLOE~\cite{KLOE+06:KSe3,KLOE+06:KSratio}).
The results for $|f^{K^0}_+(0)V_{us}|$ are averaged taking the error correlations into account~\cite{Flavia+08}. 
The measurements are compatible with each other, with an average of 
$|f^{K^0}_+(0)V_{us}|=0.2166(5)$ and a fit $\chi^2$ probability of 58\%. 

The universality of weak vector transitions dictates the equality of the 
effective coupling constants extracted for $K_{e3}$ and $K_{\mu3}$ decays. The ratio $R_{\mu e3}$ defined
as
\begin{equation}
R_{\mu e}=\frac
{\Gamma^i(K_{\mu3(\gamma)})I^{i}_{e3} (1+\delta_{e3}^{i})}
{\Gamma^i(K_{e3(\gamma)}) I^{i}_{\mu3}(1+\delta_{\mu3}^{i})},
\end{equation}
is indeed proved to be compatible with unity: $R_{\mu e}=1.0043(52)$~\cite{Flavia+08}. 
This was not the case at the time of the 2004 PDG compilation~\cite{PDG+04}:
\begin{equation}
\begin{tabular}{|l|l|l|}\hline
 Mode         & \multicolumn{2}{c|}{$R_{\mu e}$}   \\ 
              & FlaviaNet average 2007 & PDG 2004 \\ \hline
$K_{L}$       & 1.0049(61)      &  1.054(15)      \\
$K^\pm$       & 1.0029(86)      &  1.019(13)      \\ \hline
\end{tabular}
\end{equation}
These results can be compared with those from the study of $\tau\to l\nu\nu$ 
decays, which are sensitive as well to LFV effects in weak vector decays.
The world-average result from $\tau$ decays, 
$R_{\mu e}=0.9998(40)$~\cite{PDG+08}, has comparable error with that 
from $K$ decays. 
 
\begin{figure}[h]
\centering
\includegraphics[width=60mm]{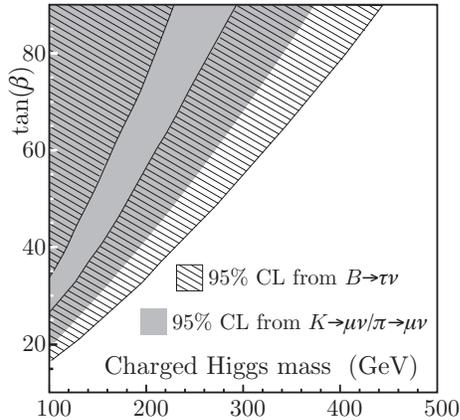}
\caption{Excluded regions at 95\% CL from analysis of decays $K\to\mu\nu$ 
(filled area) and $B\to\tau\nu$ (hatched area).} \label{fig:vusvudfit2}
\end{figure}

Measurements of $K_{l2}$ widths can be linked to new physics effects. 
The ratio of $K_{\mu2}$ to $\pi_{\mu2}$ decay widths might accept
NP contributions from charged Higgs exchange~\cite{hou,isidoriparadisi} 
in supersymmetric extensions of the SM with two Higgs doublets. 
The effect would be seen as a difference of the $V_{us}/V_{ud}$ ratio 
extracted from $K_{\mu2}$, $\pi_{\mu2}$ and that 
extracted from $K_{l3}$ and superallowed Fermi transitions (``$0^+$''):
\[
\left|\frac
{V_{us}(K_{l2})V_{ud}(0^+)}{V_{us}(K_{l3})V_{ud}(\pi_{l2})}
\right|=\left|1-\frac{m^2_{K}(m_s-m_d)\tan^2\beta}{M^2_{H}m_s(1+\epsilon_0\tan\beta)}
\right|,
\]
where $\tan\beta$ is the ratio of up- and down-Higgs vacuum expectation values, $M_{H}$ is the charged Higgs mass, and
$\epsilon_0\sim0.01$~\cite{isidoriretico}.
The world average result of this ratio 
can be translated into a 95\%-CL exclusion plot in the plane $\tan\beta$ vs $M_H$ 
(see Fig.~\ref{fig:vusvudfit2}), showing that this analysis is complementary to and competitive with 
that~\cite{isidoriparadisi} 
using the average $\mathrm{BR}(B\to\tau\nu)=1.42(44)\times10^{-4}$ of Babar and Belle measurements~\cite{bellebabar}.

A strong interest for a new measurement of the ratio
$R_K=\Gamma(\ke)/\Gamma(\km)$
has recently arisen, triggered by the work of Ref.~\cite{masiero}. 

The SM prediction of $R_K$ benefits from
cancellation of hadronic uncertainties to a large extent and therefore
can be calculated with high precision. Including radiative
corrections, the total uncertainty is less than 0.5 per
mil~\cite{ciriglianorosell07}. Since the electronic channel is
helicity-suppressed by the $V-A$ structure of the charged weak
current, $R_K$ can receive contributions from physics beyond the SM,
for example from multi-Higgs effects inducing an effective
pseudoscalar interaction.  It has been shown in Ref.~\cite{masiero}
that deviations from the SM of up to few percent on $R_K$ are quite
possible in minimal supersymmetric extensions of the SM and in
particular should be dominated by lepton-flavor violating
contributions with tauonic neutrinos emitted in the electron channel. 

In order to compare with the SM prediction at this level of accuracy,
one has to treat carefully the effect of radiative corrections, which
contribute to nearly half the $K_{e2}$ width.  In particular, the
SM prediction of Ref.~\cite{ciriglianorosell07} is made considering all
photons emitted by the process of internal bremsstrahlung (IB) while
ignoring any contribution from structure-dependent direct emission
(DE). Of course both processes contribute, so in the analysis
DE is considered as a background which can be distinguished from the IB
width by means of a different photon energy spectrum.

Two experiments are participating in the challenge to push the error on $R_K$ from the present
6\% down to less than 1\%. Last year, KLOE and NA48/2 announced preliminary 
results~\cite{KLOE+07:ke2,NA48+07:ke2} with errors ranging from 2\% to 3\%.
Moreover, the new NA62 collaboration collected more than 100\,000 $K_{e2}$ events in a 
dedicated run of the NA48 detector, aiming at reaching an accuracy of few per mil on $R_K$.
The analyses of $K_{e2}/K_{\mu2}$ in the KLOE and NA48/NA62 setups will be the main topic discussed here.

\section{Measuring $R_K$ at KLOE}
DA$\Phi$NE, the Frascati $\phi$ factory, is an $e^{+}e^{-}$ collider
working at $\sqrt{s}\sim m_{\phi} \sim 1.02$~GeV. $\phi$ mesons are produced,
essentially at rest, with a visible cross section of $\sim$~3.1~$\mu$b
and decay into $K^+K^-$ pairs with a BR of $\sim 49$\%.

Kaons get a momentum of $\sim$~100~MeV/$c$ which translates into a low speed, $\beta_{K} \sim$ 0.2.
$K^+$ and $K^-$ decay with a mean length of $\lambda_\pm\sim $~90~cm and can be 
distinguished from their decays in flight to one of the two-body final states 
$\mu\nu$ or $\pi\pi^0$.

The kaon pairs from $\phi$ decay are produced in a pure $J^{PC}=1^{--}$ quantum state, so that 
observation of a $K^+$ in an event signals, or tags, the presence of a $K^-$
and vice versa; highly pure and nearly monochromatic $K^\pm$
beams can thus be obtained and exploited to achieve high precision in the measurement of 
absolute BR's.

The analysis of kaon decays is performed with the KLOE detector, consisting essentially of a drift chamber, DCH, surrounded by an
electromagnetic calorimeter, EMC. A superconducting coil provides a 0.52~T magnetic field.
The DCH~\cite{nimdch} is a cylinder of 4~m in diameter
and 3.3~m in length, which constitutes a fiducial volume 
for $K^\pm$ decays extending for $\sim1\lambda_\pm$, respectively.
The momentum resolution for tracks 
at large polar angle is $\sigma_{p}/p \leq 0.4$\%. 
The c.m.\ momenta reconstructed from identification of 1-prong $K^\pm\to\mu\nu,\pi\pi^0$ decay vertices in the DC 
peak around the expected values with a resolution of 1--1.5~MeV, thus allowing clean and efficient $K^\mp$ tagging. 

The EMC is a lead/scintillating-fiber sampling calorimeter~\cite{nimcalo}
consisting of a barrel and two endcaps, with good
energy resolution, $\sigma_{E}/E \sim 5.7\%/\sqrt{\rm{E(GeV)}}$, and excellent 
time resolution, $\sigma_{T} =$~54~ps$/\sqrt{\rm{E(GeV)}} \oplus 50$ ps. 
The timing capabilities of the EMC are exploited to precisely reconstruct 
the position of decay vertices of $K^\pm$ to $\pi^0$'s from the
cluster times of the emitted photons, thus allowing a precise measurement of the $K^\pm$ lifetime~\cite{KLOE+08:KPlife}.

In early 2006, the KLOE experiment completed data taking, having collected
$\sim2.5$~fb$^{-1}$ of integrated luminosity at the $\phi$ peak,
corresponding to $\sim$3.6 billion $K^+K^-$ pairs.
Using the present KLOE
dataset, the KLOE collaboration claims that an accuracy of about 1~\% in the
measurement of $R_K$ might be reached.

Given the $K^\pm$ decay length of $\sim$90~cm, the selection of
one-prong $K^{\pm}$ decays in the DC required to tag $K^{\mp}$ has an
efficiency smaller than 50\%. In order to keep the statistical
uncertainty on the number of \ke counts below 1\%,
a ``direct search'' for \ke and \km decays is performed, without
tagging. Since the wanted observable is a ratio of BR's for two channels with
similar topology and kinematics, one expects to benefit from some
cancellation of the uncertainties on tracking, vertexing, and
kinematic identification efficiencies.  Small deviations in the
efficiency due to the different masses of $e$'s and $\mu$'s will be
evaluated using MC.

Selection starts by requiring a kaon track decaying in a DC fiducial
volume (FV) with laboratory momentum between 70 and 130~MeV, and a
secondary track of relatively high momentum (above 180~MeV).
The FV is defined as a cylinder parallel to the beam axis with length
of 80~cm, and inner and outer radii of 40 and 150~cm, respectively.
Quality cuts are applied to ensure good track fits.

\begin{figure}[ht]
 \center
\includegraphics[width=40mm]{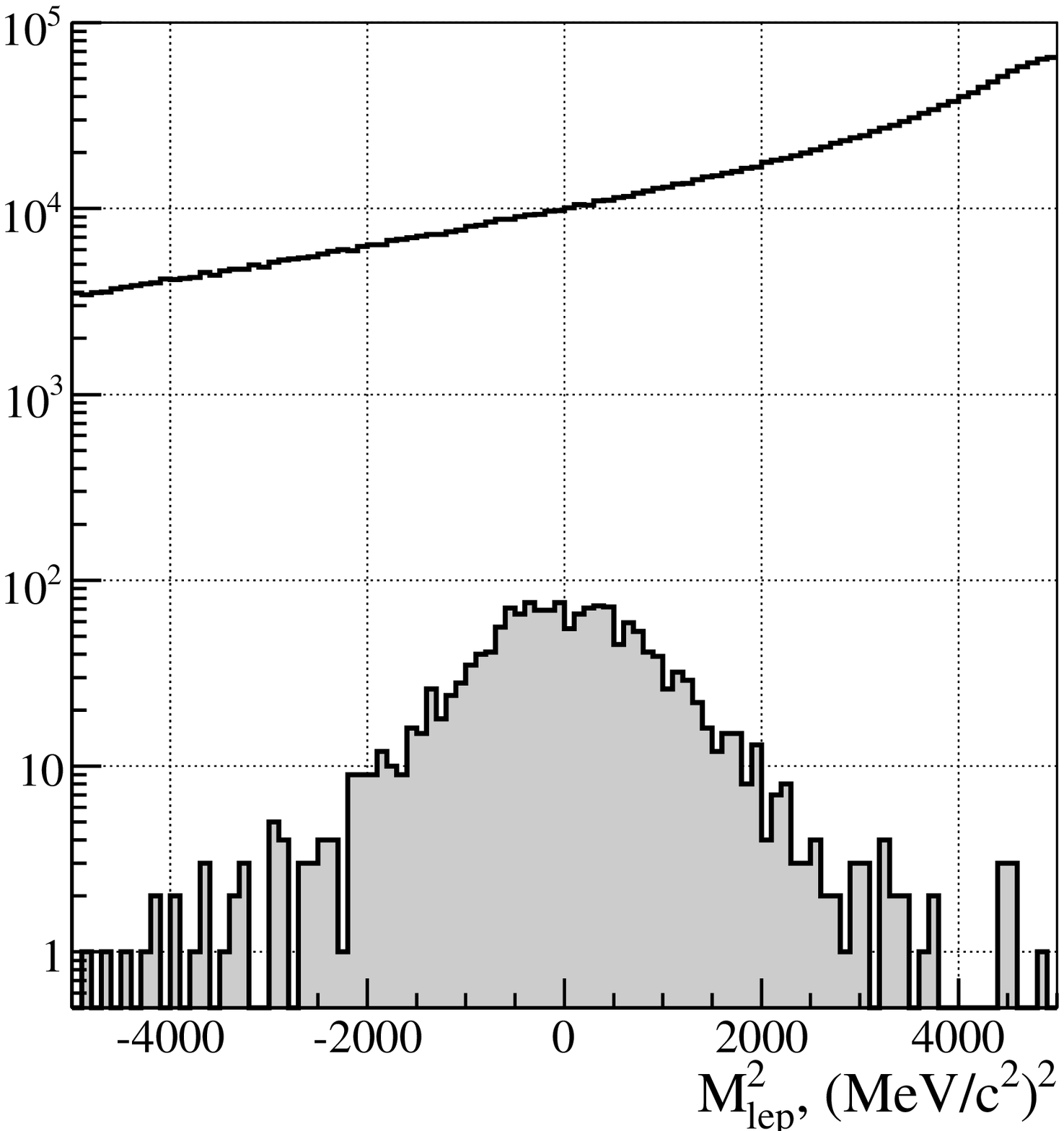}
\includegraphics[width=40mm]{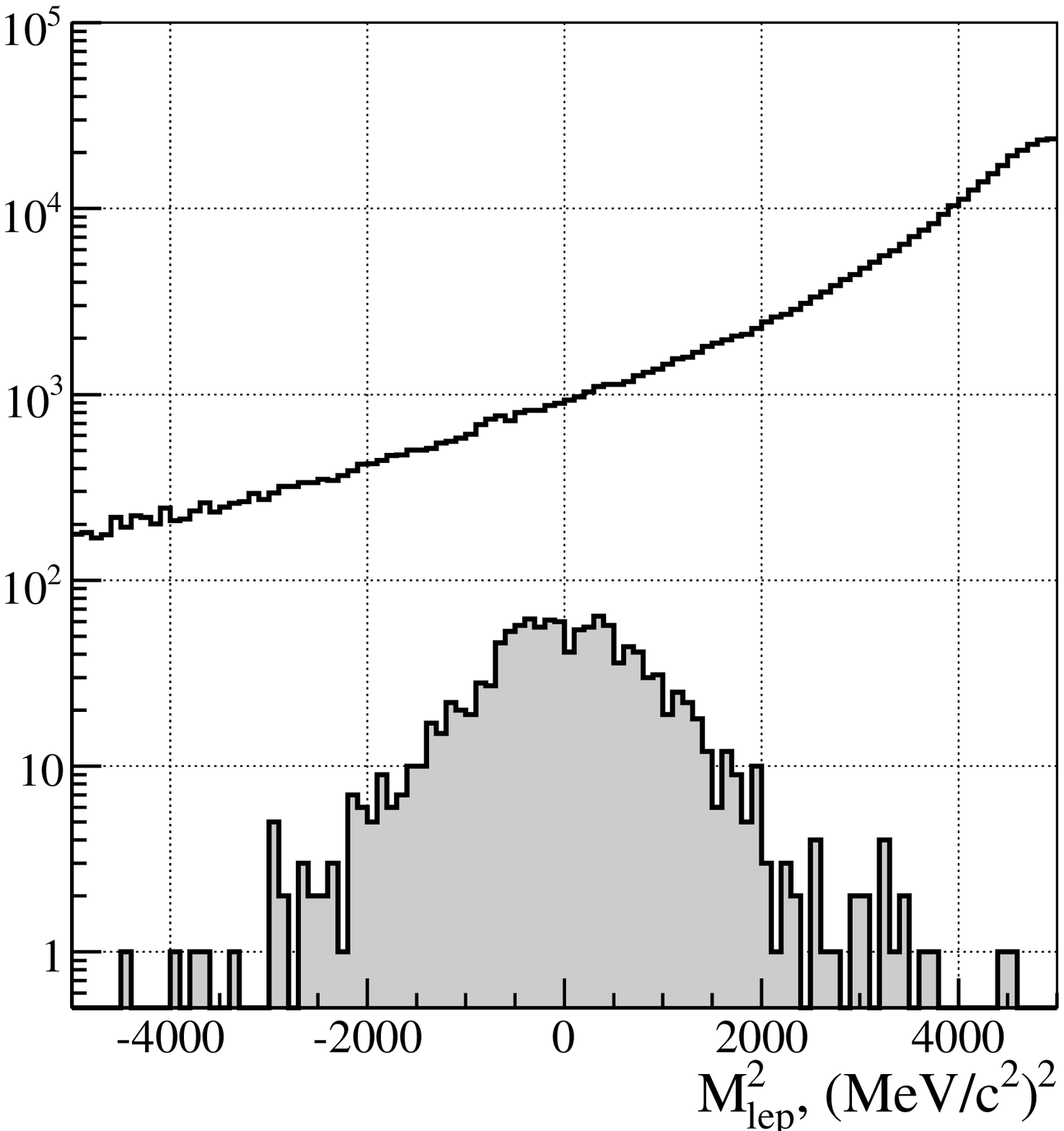}
\caption{MC distribution of $M_\mathrm{lep}^2$ before (left) and after
(right) quality cuts are applied. Shaded histogram: \ke events. Open histograms: background,
dominated by \km events. In MC, $R_K$ is set to the SM
value.}
\label{ke2:mlep}
\end{figure}

A powerful kinematic variable used to distinguish \ke and \km decays
from the background is calculated from the momenta of the kaon and the
secondary particle measured in DC: assuming zero neutrino mass one can
obtain the squared mass of the secondary particle, or lepton mass
($M_\mathrm{lep}^2$). The distribution of $M_\mathrm{lep}^2$ is shown
in Fig.~\ref{ke2:mlep} for MC events before and after quality cuts are
applied. While this selection is enough for clean
identification of a \km sample, further rejection is needed in order
to identify \ke events: the background, which is dominated by badly
reconstructed \km events, is reduced by a factor of $\sim$10 by the
quality cuts, but still remains $\sim$10 times more frequent than the
signal in the region around the electron mass peak. 

\begin{figure}[ht]
 \center
\includegraphics[width=80mm]{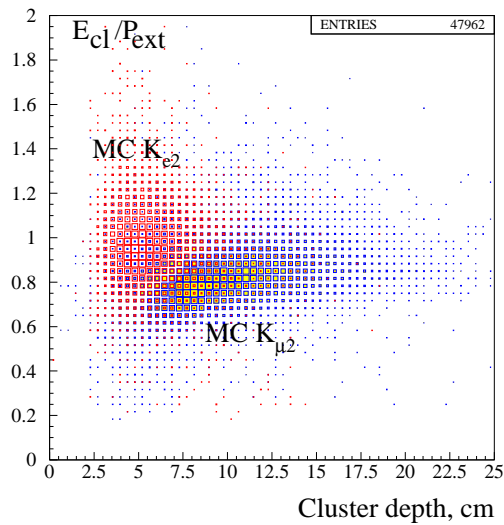}
\caption{MC distribution of the ratio $E_\mathrm{cl}/P_\mathrm{ext}$
of cluster energy and track momentum as a function of the depth of the
cluster along the direction of impact of the secondary particle on the
EMC.}
\label{ke2:ell}
\end{figure}

Information from the EMC is used to improve background rejection. The
secondary track is extrapolated to a position $\vec{r}_\mathrm{ext}$
on the EMC surface with momentum $\vec{p}_\mathrm{ext}$ and associated
to the nearest calorimeter cluster satisfying the impact-parameter cut
$d_\perp<30$~cm, where $d_\perp =
|\vec{p}_\mathrm{ext}/|p_\mathrm{ext}|\times(\vec{r}_\mathrm{ext}-\vec{r}_\mathrm{cl})|$.
For electrons, the associated cluster is close to the EMC surface so
that its position projected along the track $d_\parallel=
|\vec{p}_\mathrm{ext}\cdot(\vec{r}_\mathrm{ext}-\vec{r}_\mathrm{cl})|$
is only a few cm. Moreover, for electrons the cluster energy
$E_\mathrm{cl}$ is a measurement of the particle momentum
$p_\mathrm{ext}$. Therefore the following condition is required in the
plane $E_\mathrm{cl}/p_\mathrm{ext}$ vs $d_\parallel$ (see Fig.~\ref{ke2:ell}):
\begin{equation}
\label{eq:elli}
\left(\frac{d_\parallel[\mathrm{cm}]-2.6}{2.6}\right)^2+
\left(\frac{E_\mathrm{cl}/p_\mathrm{ext}-0.94}{0.2}\right)^2<2.5.
\end{equation}

Electron clusters can be further distinguished from $\mu$ (or $\pi$)
clusters by exploiting the granularity of the EMC: electrons shower
deposits their energy mainly in the first plane of EMC, while muons
behave like minimum ionizing particles in the first plane and they
deposit a sizable fraction of their kinetic energy from the third
plane onward when they are slowed down to rest (Bragg peak). 

The
particle identification (PID) was therefore based on the asymmetry
$A_f$ of energy deposits between the second and the first planes hit,
on the spread $E_\mathrm{RMS}$ of energy deposits on each plane, on
the position $r_\mathrm{max}$ of the plane with the maximum energy,
and on the asymmetry $A_l$ of energy deposits between the last and the
next-to-last planes.  Muon clusters with the signature $A_f>0$, or
$x_\mathrm{max}>12$~cm, or $A_l<-0.85$ are rejected. This criteria
were optimized during MC and control sample studies. 

The PID technique described above selects \ke events with an
efficiency $\epsilon^\mathrm{PID}_{Ke2}\sim64.7(6)\%$ and a rejection
power for background of $\sim300$. These numbers have been evaluated
from MC. 

\begin{figure}[ht]
 \center
\includegraphics[width=40mm]{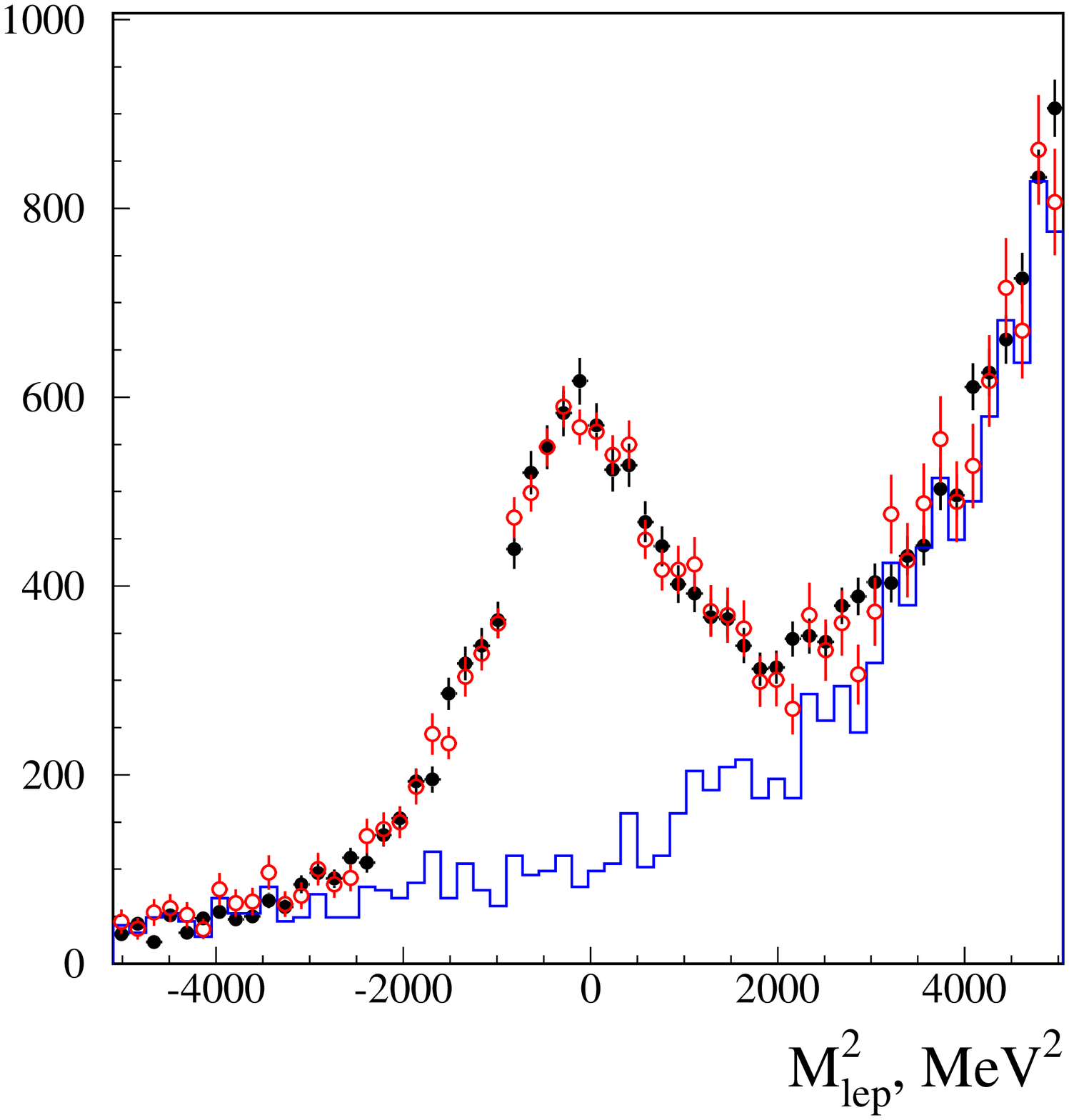}
\includegraphics[width=40mm]{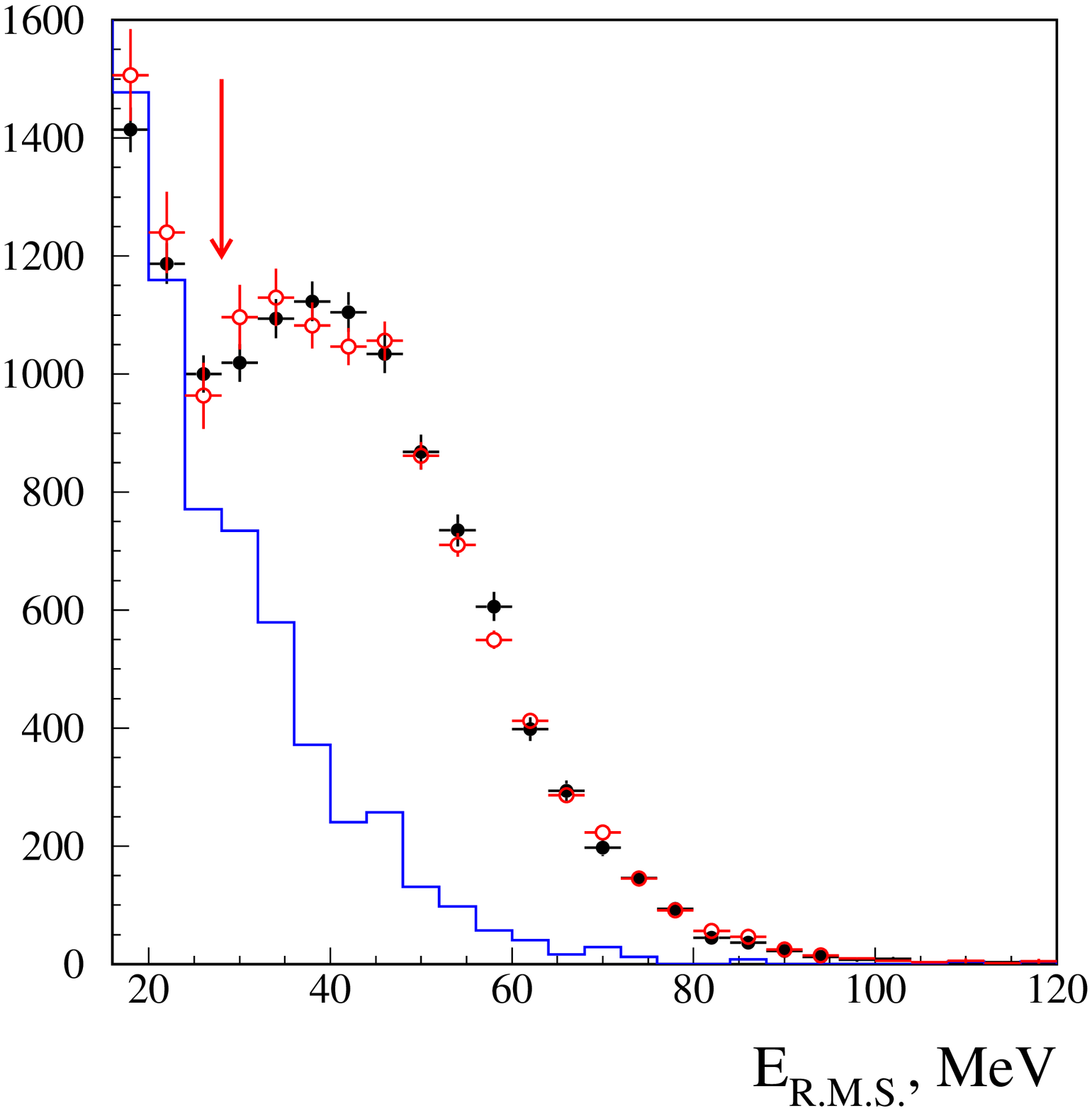}
\caption{Distributions of the lepton mass squared $E^2_\mathrm{lep}$
of the secondary track (left panel) and of the spread $E_\mathrm{RMS}$
of the energy deposits among the planes of the connected cluster in
the EMC (right panel). Filled dots represent the data, open dots
are the result of a maximum-likelihood fit using signal and
background (solid line) distributions as input.}
\label{ke2:fit}
\end{figure}

A likelihood fit to the two-dimensional $E_\mathrm{RMS}$ vs
$M_\mathrm{lep}^2$ distribution was performed to get the number of
signal events. Distribution shapes for signal and background were
taken from MC; the normalizations for the two components are the only
fit parameters.  The number of signal events obtained from the fit is
$N_{Ke2}=8090\pm156$. Projections of the fit results onto the two axes
are compared to real data in Fig.~\ref{ke2:fit}.

\section{Measuring $R_K$ at NA48 and NA62}

Unseparated, simultaneous, and highly collimated $K^+$ and $K^-$ beams were designed to precisely
measure the asymmetry in the dalitz-plot density for the decays $K^\pm\to\pi^\pm\pi^{+,0}\pi^{-,0}$ 
with the NA48/2 experiment.

The beams were produced by primary, 400-GeV protons from the SPS beam interacting with a beryllium target.
After passing a selection system made of four dipole magnets with zero total deflection (``achromat'') 
and various collimators, kaons enter a 114-m long, $\sim$2-m in diameter, cylindrical 
fiducial decay volume $\sim180$~m after they are produced from the target, with a 
momentum of $\sim60$~GeV and a momentum bite of $\sim3.8\%$. The NA48 detector is described in Ref.~\cite{NA48+07:detector}.
Kaon decay products are tracked by a magnetic
spectrometer consisting of two pairs of drift chambers (DCHS) interleaved 
with a dipole magnet. The dipole provides a transverse momentum kick
of $\sim121$~MeV. The relative momentum resolution from the tracking system is 
$\sigma_p/p\sim1.02\%\oplus0.044\%/p[\mathrm{GeV}]$.

Particles emerging from the last drift chamber reach a hodoscope system made of scintillator slabs running along
orthogonal views. This system is used for
initiating the trigger and establishing the event time: the time resolution is $\sim$~150~ps. 

An e.m. calorimeter made of a LKr-filled volume closes the decay volume $\sim 240$~m after the target. 
The calorimeter is 
read out by a fine-granularity system of accordion-shaped cathode rods providing
highly uniform and efficient vetoing for photons impacting its $80\times 80~\mathrm{cm}^2$ surface, 
and guaranteeing excellent energy and transverse position resolutions for e.m. showers: 
\begin{eqnarray}
\sigma_E/E=3.2\%/\sqrt{E[\mathrm{GeV}]}\oplus9\%/E[\mathrm{GeV}]\oplus0.42\%\\
\sigma_{x,y}=4.2\mathrm{mm}/\sqrt{E[\mathrm{GeV}]}\oplus0.6\mathrm{mm}.
\end{eqnarray}
For the $K_{e2}/K_{\mu2}$ analyses, the starting samples are ``one-track'' triggers obtained by requiring
hits in both views of the hodoscope. While in 2003 a downscaled unbiased sample of one-track events was
obtained with a 12-hour run, in 2004 a 56-hour special run has been made with different triggers for $K_{e2}$ and
$K_{\mu2}$, the condition of having more than 10~GeV deposited in the LKr calorimeter being added for the
first channel. 

\begin{figure}[ht]
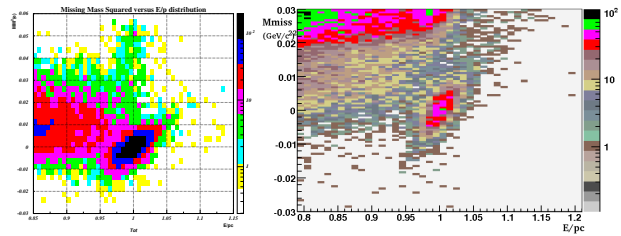

 \center
\rotatebox{270}{\includegraphics[width=30mm]{signal_fiorini_phdthesis.epsi}}
\rotatebox{270}{\includegraphics[width=30mm]{signal_venelin_kaon2005.epsi}}
\caption{Distributions of the missing mass squared 
of the secondary track in the electron mass hypothesis versus the ratio $E/P$ of cluster energy in the LKr
and track momentum. Left (right) panel refers to 2003 (2004) data.}
\label{ke2na48-2:esupvsmmiss}
\end{figure}

Data reduction for both samples is made by requiring only one track reconstructed by DCHS passing acceptance and
quality cuts; the track must intercept the beam line with a small impact parameter in a point well
within the decay volume, between 2~m and 8.5~m from the last collimator; 
the track momentum is required to be in the expected range for $K_{l2}$ decays, $15<p<50$~GeV.
These cuts reject much of the background from non-kaon events or early kaon decays. 
Background rejection for $K_{e2}$ identification is performed by associating the track from DCHS 
with a LKr cluster: the ratio $E/P$ of cluster energy and track momentum plotted for 2003 and 2004 data 
versus the missing mass (evaluated from track momentum) shows a clear peak corresponding to $K_{e2}$ decays,
see Fig.~\ref{ke2na48-2:esupvsmmiss}. 

\begin{figure}[ht]
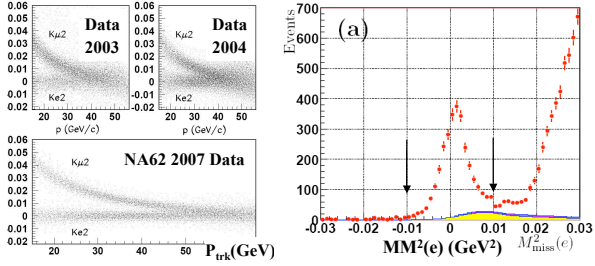

\center
\rotatebox{270}{\includegraphics[width=35mm]{mmiss_ptrk_years.epsi}}
\rotatebox{270}{\includegraphics[width=35mm]{mmiss_ke2peak_na62.epsi}}
\caption{Left: missing mass squared $MM^2(e)$ (GeV$^2$ units), evaluated in the electron mass hypothesis, 
as a function of the track momentum; different years of data taking are separately shown. Right: $K_{e2}$ peak in the
$MM^2(e)$ distribution for the 2007 NA62 run; a background estimate is also shown (filled histogram).}
\label{ke2naxx:mmiss}
\end{figure}

The main background under the $K_{e2}$ peak is due to 
$K_{\mu2}$ events in which the muon
produces a ``catastrophic'' energy release in the LKr, with $E/P\sim1$. A cut on $M_\mathrm{miss}$ is
enough for clean identification of $K_{e2}$ for low momentum tracks, namely $p<\sim30$~GeV, 
while kinematics cannot be used at high momenta; see left panel of Fig.~\ref{ke2naxx:mmiss}. 

For the purpose of background subtraction in this range, the probability 
for a muon to produce a cluster with $E/P>0.95$ has been evaluated from a selected muon sample and is
$\sim0.5\times10^{-6}$. The statistics of the sample used induces
a significant error to the final result: for the preliminary result from 2004 data, the fractional statistical
error due to $K_{e2}$ counts is 1.85\%, while the systematic error due to background subtraction is 1.59\%.

\begin{figure}[ht]
\center
    \includegraphics[width=40mm]{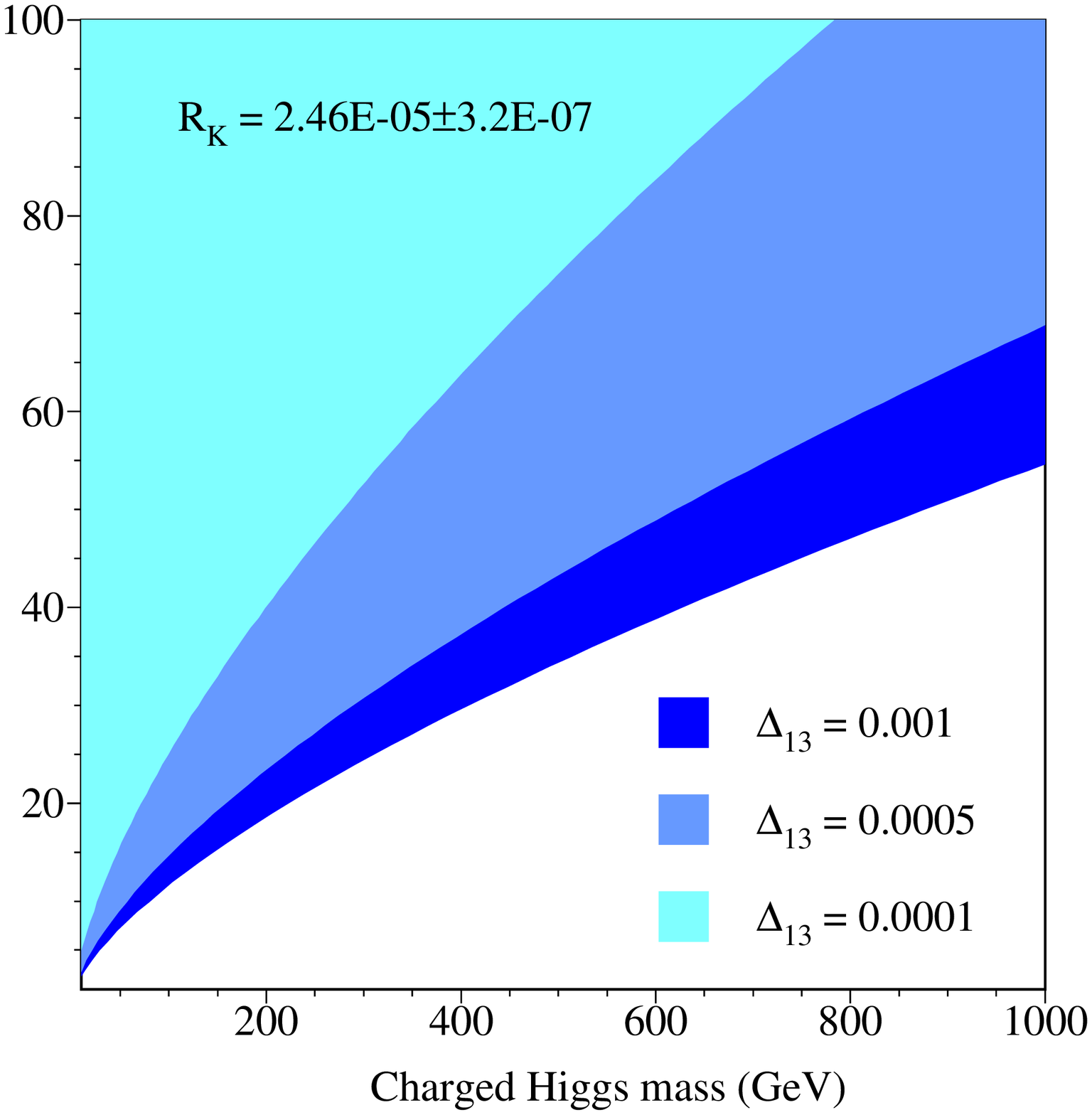}
    \includegraphics[width=40mm]{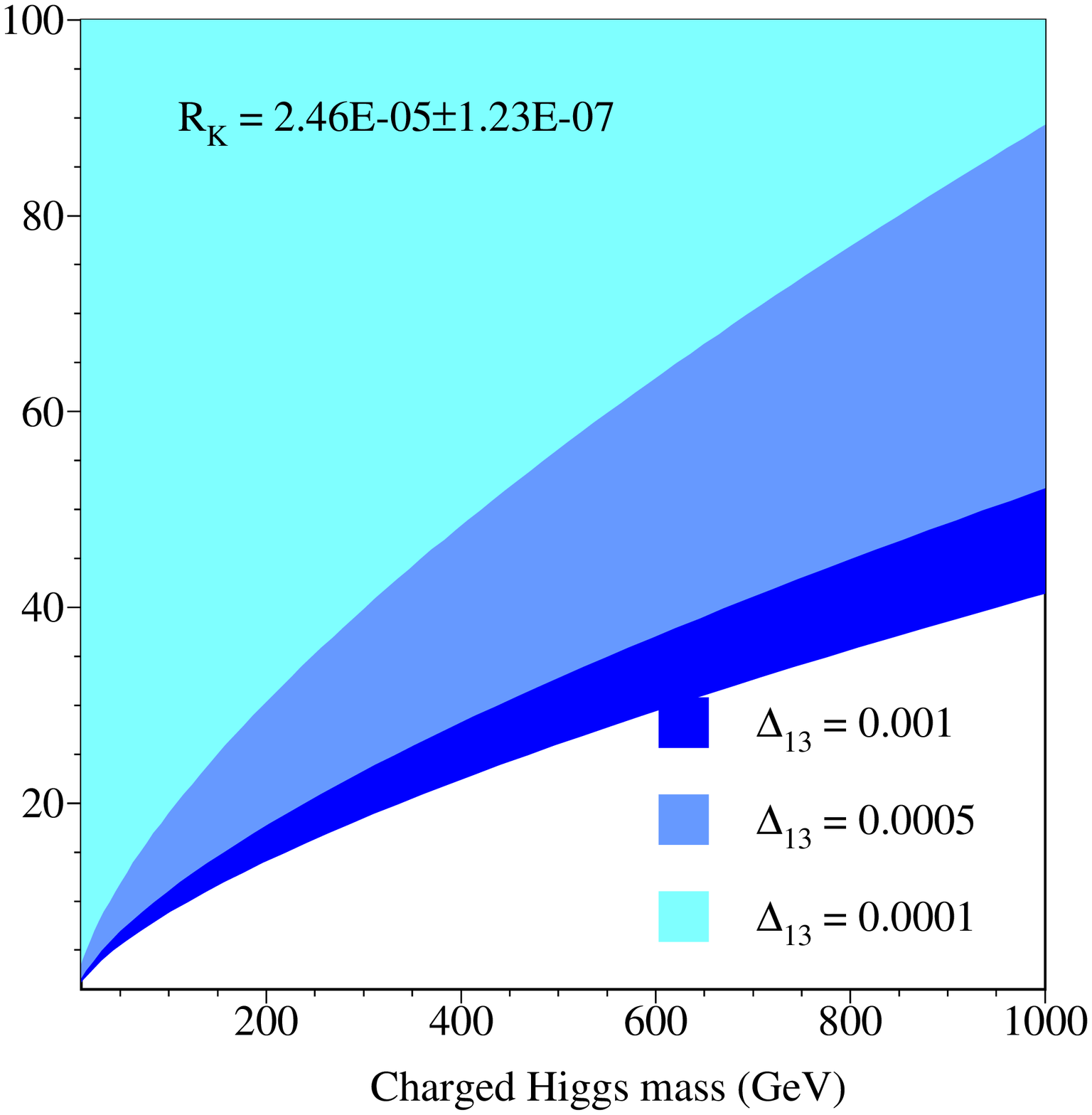}
\caption{Exclusion plots at 95\% CL in the $\tan\beta$-$M_H$ plane from the present world average (left) or from a result with the same central value but a 0.5\% error (right).
Different values of the $\Delta_{13}$ LFV parameter of Ref.~\cite{masiero} are used.}
\label{fig:rkplots}
\end{figure}

For these reasons, the experimental setup has been optimized by the NA62 Collaboration for the purpose of a
new dedicated data taking during 2007: the average kaon momentum selected has been raised up to 75~GeV, 
the momentum bite has been lowered to 2.5\%, and the transverse momentum kick induced by the analyzing 
magnet of the DCHS system has been increased
by more than a factor of 2, up to 263~MeV. These features improve the $M_\mathrm{miss}$ resolution and maximize
the kinematical separation of $K_{e2}$ and $K_{\mu2}$, see right panel of Fig.~\ref{ke2naxx:mmiss}. 
For the evaluation of the probability for a 
muon to fake an electron in the LKr calorimeter, for part of the run a lead wall has been put in between 
the two hodoscope scintillator layers:
this allows a direct measurement of the ``catastrophic'' energy loss probability from data. The data taking
lasted for 4 months, and allowed the NA62 Collaboration to acquire the largest $K_{e2}$ sample in the world,
amounting to more than $10^5$ events; see right panel of Fig.~\ref{ke2naxx:mmiss}. The analysis is expected to measure
$R_K$ with a $<0.5\%$ total error, thus improving significantly the sensitivity to new physics contributions
(Fig.~\ref{fig:rkplots}).

\end{document}